\documentclass[twocolumn,final,showpacs,
               showkeys,superscriptaddress,
               aps,superbib,
               letterpaper,balancelastpage,
               lengthcheck,byrevtex,bibnotes,
               footinbib]{revtex4}
\usepackage{graphics}
\usepackage{amsmath}
\usepackage{amssymb}
\usepackage{amsfonts}
\usepackage{graphicx}
\usepackage[squaren]{SIunits}

\def\ie{{\em i.e.}}

\def\etc{{\em etc.}}

\def\sizeFIG{.9}
\def\sizeFIGa{.9}

\def\kesik{\protect\mbox{--\, --\, --}}
\def\chain{\protect\mbox{-- $\cdot$ --}}
\def\dd{{\rm d}}
\def\lila{$\textrm{Li}_{0.5}\textrm{La}_{0.5}\textrm{TiO}_{3}$}

\begin{document}

\title{Comparison of methods for estimating continuous distributions of relaxation times}
\author{Enis Tuncer}
\email[Email address: ]{enis.tuncer@physics.org}
\altaffiliation[Present address: ]{Applied Superconductivity Group, High Voltage and Dielectrics Division, Oak Ridge National Laboratory, Oak Ridge, 37831 TN, USA}
\affiliation{Solid State Physics Division, {\AA}ngstr{\"o}m Laboratory, Uppsala University, SE-75121 Uppsala, Sweden}
\author{\firstname J. Ross \surname Macdonald}
\email[Email address: ]{macd@email.unc.edu}
\affiliation{Department of Physics and Astronomy, University of North Carolina, Chapel Hill, NC 27599-3255, USA}
\begin{abstract}
{The nonparametric estimation of the distribution of relaxation times approach is not as frequently used in the analysis of dispersed response of dielectric or conductive materials as are other immittance data analysis methods based on parametric curve fitting techniques. Nevertheless, such distributions can yield important information about the physical processes present in measured material. In this letter, we apply two quite different numerical inversion methods to estimate the distribution of relaxation times for glassy \lila\ dielectric frequency-response data at $225\ \kelvin$. Both methods yield unique distributions that agree very closely with the actual exact one accurately calculated from the corrected bulk-dispersion Kohlrausch model established independently by means of parametric data fit using the corrected modulus formalism method. The obtained distributions are also greatly superior to those estimated using approximate functions equations given in the literature.  
}
\end{abstract}
\keywords{Distribution of relaxation times; dielectric relaxation; least squares approximations; Monte Carlo methods; inverse problems}
\pacs{77.22.Gm 02.70.Rr 02.30.Zz 02.50.Ng 02.60.-x 02.30.Sa 66.30Dn 61.47.Fs 72.20.i 66.10.Ed}

\maketitle

Broadband dielectric (also known as immittance or impedance) spectroscopy is widely used to characterize materials and to help understand the mechanisms involved in such challenging areas of condensed-matter physics as conductivity, molecular relaxation, liquid-glass transition \etc\ 
\cite{Jonscher1983}. In this experimental technique an electrical property of the material is recorded as a function of probing field frequency $\nu$. Data may be expressed at one of the four specific immittance levels ({\em i}) the complex resistivity $\rho(\omega)$; ({\em ii}) the complex modulus $M(\omega)\equiv\imath\omega\varepsilon_0\rho(\omega)$; ({\em iii}) the complex permittivity $\varepsilon\equiv[M(\omega)]^{-1}$; and ({\em iv}) the complex conductivity $\sigma(\omega)\equiv\imath\omega\varepsilon_0\varepsilon(\omega)\equiv[\rho(\omega)]^{-1}$. Here, $\omega$ is the angular frequency $\omega=2\pi\nu$; $\varepsilon_0$ is the permittivity of free space; and $\imath=\sqrt{-1}$. 

Once a data set is acquired, it may be expressed at an appropriate immittance level and then analyzed to obtain valuable information about material processes. Often employed procedures that have been used to  analyze frequency response data are (a) using the Kohlrausch-Williams-Watt (KWW) approached derived from stretched exponential behavior in the time domain~\cite{Kohl,
McCrum}; (b) the Havriliak-Negami empirical expression~\cite{HN}; and (c) estimation of the distribution of relaxation times (DRT) inherent in the data~\cite{BottcherDRT,RossBJP,McCrum,
macdonald:6241,TuncerLicDRT,Dias}
, an approach not as commonly employed as the other two. Unlike the KWW analysis of (a), procedure (b) is a data fitting method that does not lead to added understanding of the physical processes presented in the experimental material. On the other hand, KWW analyses involve fitting models whose parameters are all of physical significance. Although they are useful for comparing fit parameters for various materials at different state variable levels they are less appropriate for data involving several DRTs associated with different physical processes. 

The DRT approach of (c) is an elegant method for investigating the contributions of relaxing units to  the total relaxation and for determining the influence of state variables on the relaxation. In the presence of different processes or broad relaxations, the DRT approach is superior to the parametric ones since (1) no {\em a priori} assumptions are needed, \ie, a sum of empirical expressions \etc; 
(2) the actual distributions in a given data set are initially unknown; (3) a DRT can be related to various physical parameters of the system; (4) and when there are two different overlapping relaxations present, their depencies on state variables would be easy to identify and to observe the influence of the state variables on the distributions. 
As an example, the dynamic complexity of the relaxation system can be determined by estimating its DRT and thus establishing whether it is intrinsicly broadening (homogeneous) or a distribution of responses (heterogeneous)~\cite{Stoneham1969}. A distribution may be characterized as either discrete (composed of individual points) or continuous, and DRT analysis can unambiguously distinguish between these two possibilities~\cite{macdonald:6241,TuncerLicDRT}. 
Recently, non-resonant spectral hole burning technique has been employed to resolve distinct continuous distributions experimentally in order to identify multiple relaxing domains in materials~\cite{BoehmerScience}.
\begin{figure}[t]
  \centering
  \includegraphics[width=\sizeFIGa\linewidth]{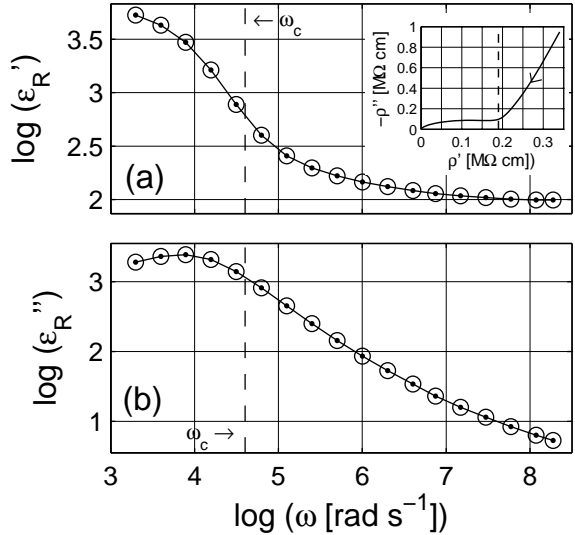}  
  \caption{(a) Real and (b) imaginary parts of the raw dielectric permittivity of LLT at 225 K. The solid line and solid points shows the full data, including electrode effects and that associated with ohmic conductivity. The open-circle symbols show the predictions of a full complex nonlinear least-squares fit of this data set, as described in the text. The subscript 'R' in the axis labels indicates that the data are presented without any transformation (raw data).  The inset shows the same data plotted at the complex resistivity level, and the arrow in the inset indicates the direction of increasing frequency.  The vertical dashed line shows the division between the two types of response present and defines the critical radial frequency $\omega_c$.}
  \label{fig:Fig1}
\end{figure}

In this letter, we compare the results of two different DRT inversion methods for analyzing a set of experimental frequency-response data that involves a continuous distribution. We also compare the accuracy of two equations for estimating appropriate distribution functions proposed by \citet{BottcherDRT}. Although estimation of discrete-point distributions is not an ill-posed mathematical problem~\cite{macdonald:6241}, distribution estimation of continuous distributions, the usual situation, is ill-posed. It is therefore particularly important to assess the utility and power of different DRT estimation procedures for a well-defined data situation.

Experimental data, with $M=52$ points, for the \lila\ (LLT) glass 
at $225\ \kelvin$~\cite{LeonJNCS}, expressed as the complex resistivity and dielectric levels, were found to involve an appreciable component associated with  electrode polarization effects. LLT conducts by ionic hopping and involves a finite dc resistivity, $\sigma_0\equiv\sigma(0)$. Further analysis of data for this material over a range of temperatures established that both $\sigma_0$ and the characteristic relaxation time of the dispersion of the bulk material, $\tau_o$, were thermally activated with $T\sigma_0$ and $\tau_o$ having the same activation energies~\cite{Macdonald2004PRB}. 

Such behavior indicates that it is most appropriate to identify the bulk dispersive response of this material with a conductive-system dispersion of resistivity relaxation times, rather than with a dielectric-system distribution of permittivity relaxation times, one where $\sigma_0$ would be naturally interpreted as a leakage conductivity unrelated to the bulk dielectric dispersion process. Since conductive-system response has already been analyzed for this data set~\cite{Macdonald2004PRB}, and since it has been shown by data fitting that it may often be difficult to discriminate between fits of conductive-system and dielectric-system models when only a single data set is available\cite{RossBJP}, we have elected to compare the two different DRT estimation procedures by determining their dielectric-system DRTs from the present data expressed at the complex permittivity level. 
\begin{figure}[t]
  \centering
  \includegraphics[width=\sizeFIGa\linewidth]{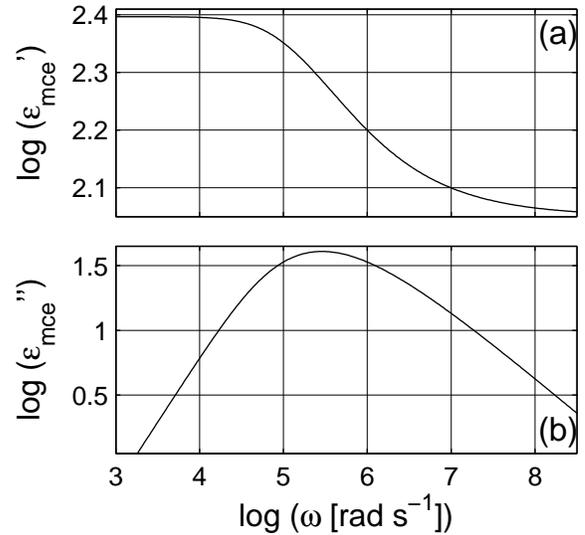}  
  \caption{(a) Real and (b) imaginary parts of the dielectric permittivity of LLT at 225 K after elimination of the contributions of ohmic conductivity and electrode polarization effects and generation of virtually exact KD-model data using LEVM.  The axis identifiers are shown with a subscript 'mce' denoting subtracted ohmic conductivity and electrode effects.}
  \label{fig:Fig2}
\end{figure}

The two analysis methods considered here will be designated \rm{I} and \rm{II}. Method \rm{I} involves a weighted nonlinear least squares approach for estimating dielectric distribution strength points, ${\sf g}_i$, at corresponding relaxation-time values $\tau_i$, with $1<i<N$~\cite{macdonald:6241}. 
It allows either discrete or continuous DRTs to be estimated in terms of the $\{{\sf g}_i,\,\tau_i\}$ values and their uncertainties, with the set of $\tau_i$'s either taken fixed or free to vary.  Better results are nearly always obtained with $\tau_i$'s taken free, as in the present work. In addition, the data may be in temporal response form or in the frequency domain involving complex response or either its real or imaginary part. An extensive fitting and inversion program named LEVM that includes Method \rm{I} is available for free downloads\cite{LEVM}. Method \rm{II} is based  on a constrained least-squares with the Monte Carlo procedure~\cite{TuncerLicDRT}. 
It leads to delta sequence distributions~\cite{Butkov} when applied to discrete DRTs~\cite{TuncerLicDRT}. 
Recently, a method rather similar to that of \rm{II} has been independently proposed, one that uses nonparametric Bayesian statistics for solving similar inversion problems~\cite{Wolpert}.

Since we are interested in the dielectric DRT for the dispersive bulk relaxation process, it is important to eliminate the contributions to the data arising from partly blocking electrode effects before estimating the DRT. To do so, a KWW response model, the KD, involving a stretched-exponential shape parameter $\beta_D$, a characteristic relaxation time $\tau_o$, and a $\Delta\epsilon$ strength parameter, was used  for fitting the original full data with inclusion of free parameters to model the electrode effects, the high-frequency-limiting bulk dielectric permittivity, $\varepsilon_{\infty}$, and $\sigma_0$~\cite{RossSSI2002}. 
The fit was excellent and yielded the following rounded estimates for $\beta_D$, $\tau_o$, $\Delta\varepsilon$, and  $\varepsilon_{\infty}$: $0.547$, $2.63\ \micro\,\second$, $137$, and $112$, respectively. 

The precise fit values of these parameters were then used in LEVM, omitting those of $\sigma_0$ and the electrode ones, to generate a set of $M=300$ data points representing just the bulk part of the Kohlrausch response to eight significant figures or better. This data set is used below to estimate its DRT by the methods mentioned above. In addition, given only values of $\beta_D$, $\Delta\varepsilon$, and a set of logarithmically distributed values of $\tau$ for the range from about $0.96\ \nano\,\second$ to $10\ \milli\,\second$, LEVM was employed to calculated highly accurate values for the KD DRT comparison with the inversion estimate.

\begin{figure}[t]
  \centering
  \includegraphics[width=\sizeFIG\linewidth]{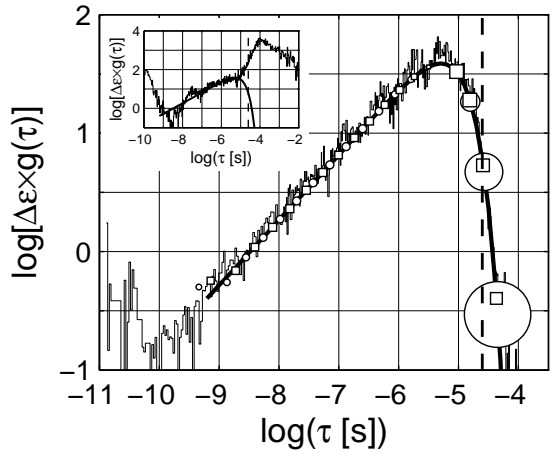}  
  \caption{Various estimates of unnormalized distribution of relaxation times following from the data presented in Fig.~\ref{fig:Fig2}.  Here the implicit scaling factor for these distributions was $\Delta\varepsilon$. The thick solid line is the unnormalized KD-model distribution. The method \rm{I} open-square symbols show points estimated using the real part of that data and the open-circle ones show those obtained from the imaginary part.  The sizes of the symbols were determined by the estimated uncertainties of the fits.  The many solid-stair lines show the results of the method \rm{II} estimation procedure using the full complex data.  The vertical dashed line here and in Fig. 4 shows the critical time constant.   For comparison, the inset presents  method \rm{II} estimates using the raw data of Fig.~\ref{fig:Fig1}.}
  \label{fig:Fig3}
\end{figure}
The complex dielectric permittivity may be expressed in terms of a general DRT formalism,
\begin{eqnarray}
  \label{eq:epsDRT}
  \varepsilon(\omega)&=&\varepsilon_\infty+(\varepsilon_s-\varepsilon_\infty)\int_{-\infty}^{\infty}\frac{{\sf g}(\ln\tau)\dd\ln\tau}{1+\imath\omega\tau}
\end{eqnarray}
where, $\varepsilon_\infty\equiv\varepsilon'(\infty)$ and $\varepsilon_s\equiv\varepsilon'(0)$ (the quantity $\Delta\varepsilon\equiv\varepsilon_s-\varepsilon_\infty$ is defined as the dielectric strength), and ${\sf g}(\ln \tau)$ is the distribution function. For a delta sequence distribution~\cite{Butkov} Eq.~(\ref{eq:epsDRT}) leads to simple Debye response~\cite{Debye1945}. Both applied methods \rm{I}  and \rm{II} are based on Eq.~(\ref{eq:epsDRT}) and are further described in Ref.~
\citet{macdonald:6241} 
and Ref.~\citet{TuncerLicDRT}
, respectively.  

In Fig.~\ref{fig:Fig1}, the complex dielectric permittivity raw data for LLT 
are presented without transformation. As evident in the inset of Fig.\ref{fig:Fig1}a the data include two different processes, with the right spur part representing low-frequency electrode polarization effects. The dashed vertical line (\kesik) in the inset indicates the approximate crossover position (shown at $40\ \kilo\,\rad\reciprocal\second$) from bulk dielectric system dispersion to conductivity and double-layer effects~\cite{RossSSI2002}. 
Since all the open-circle fit points in the figure enclose their corresponding solid data points symmetrically, one may conclude that the fit is excellent. 
\begin{figure}[t]
  \centering
  \includegraphics[width=\sizeFIG\linewidth]{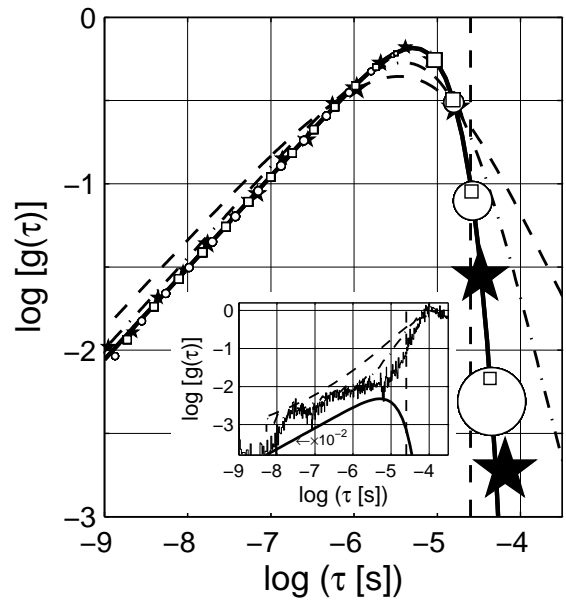}  
  \caption{Comparison of several different normalized DRT estimates. Identifications are the same as those in Fig. 3 for similar quantities.  Shown are the exact KD-model DRT, the method \rm{I} real ($\Box$) and imaginary ($\bigcirc$) data estimates,  the exponential smoothing of method \rm{II} results ($\bigstar$), and approximate DRT estimates calculated directly from the data of Fig. 2 using Eq. (2) (\kesik) and Eq. (3) (\chain).  The inset shows method \rm{II} estimates using the raw data of Fig.~\ref{fig:Fig1} compared with the scaled exact DRT (all divided by a factor of 100).} 
  \label{fig:Fig4}
\end{figure}

After we remove the contributions of the ohmic conductivity and electrode effects to the raw data, as described above, the pure dielectric-system dispersion is obtained and is presented in Fig.~\ref{fig:Fig2} and denoted by $\varepsilon_{\sf mce}$. This data set, implicitly involving the KD-model DRT, was next used to estimate the DRT by the inversion methods \rm{I} and \rm{II}. Some of these results are shown in Fig.~\ref{fig:Fig3} and \ref{fig:Fig4}. The thick solid line is that of the KD DRT with $\beta_D=0.54657$. Note that the data of Fig.~\ref{fig:Fig2} contain neither systematic nor random errors and thus allow comparison of the utility of methods \rm{I} and \rm{II} without such confounding factors. 
It is striking that the two methods both yield very accurate estimates of the exact DRT. The precision of the estimates obtained by method \rm{I} using the real part of the data (Fig.~\ref{fig:Fig3}) is the best of the results shown and is remarkably small, especially for the points at and to the left of the peak. We should also remember that increasing the number of randomly selected $\tau$ values used in method \rm{II} improves the DRT estimates; $\sim25000$ $\tau$ values were used in the present work. 

Method \rm{II} selects random $\tau$ values over a wider range than those defined by the range of the original frequency window. The range of the original data ($\varepsilon_{\sf R}$) is about $2\ \kilo\,\rad\reciprocal\second <\omega<200\ \mega\,\rad\reciprocal\second$ corresponding to $5\ \nano\,\second<\tau<500\ \micro\,\second$, somewhat smaller than the $\tau$ range following from the exact data of Fig.~\ref{fig:Fig2}, as defined above. 

In order to illustrate the utility of method \rm{II}, its DRT determined for the raw $\varepsilon_{\sf R}$ data  is shown in the inset of Fig.~\ref{fig:Fig3} with solid vertical lines and is compared to the actual distribution. Note that the presence of electrode effects results in an added distribution with a peak at $\tau=100\ \micro\,\second$. In addition, the distributions obtained from the 
$\varepsilon_{\sf R}$ and $\varepsilon_{\sf mce}$ data sets are nearly the same for $\tau<10\ \micro\,\second$ except for the presence of a small peak of the $\varepsilon_{\sf R}$ distribution estimate near $\tau\sim32\ \nano\,\second$. This could possibly be due to the raw data where no {\em a priori} assumption is made of the presence of KD-model response ($\varepsilon_{\sf mce}$). 
Also note the effects of the relaxation-time cutoff for fast processes at $\tau<1\ \pico\second$. 

To further emphasize the utility of the numerical inversion methods for estimating a DRT, we compare our results with those of \citet{BottcherDRT} in Fig.~\ref{fig:Fig4}. They derived approximate DRT expressions from the real and imaginary parts of the dielectric permittivity and their derivatives with respect to natural logarithm of angular frequency ($\ln \omega$). Two such approximation distribution functions ${\sf g}$ are listed below.
\begin{eqnarray}
  {\sf g}_{1}(\ln \omega^{-1})&=&{2\,\varepsilon''(\omega)}({\pi\,
\Delta\varepsilon})^{-1},  \label{BoetEq.9.24} \\ 
 {\sf g}_{2}(\ln \omega^{-1}) &=&-{\Delta\varepsilon}^{-1}\,{\textrm{d}\varepsilon'(\omega)}/{\textrm{d}\ln\omega}, \label{BoetEq.9.39}
\end{eqnarray}  
where the Fig.~\ref{fig:Fig1} data fit result for $\Delta\varepsilon$, $136.93$, is used along with $\varepsilon_{\sf mce}$ data set values. Clearly these expressions lead to broader distributions and to far less accurate DRT estimates than our inversion ones. In the inset of Fig.~\ref{fig:Fig4} method \rm{II} DRT estimates obtained from the raw data are again illustrated, together with those following from the application of Eqs.~(\ref{BoetEq.9.24}) and (\ref{BoetEq.9.39}). 


In conclusion, two approaches for estimating DRT in conductive and dielectric systems are applied to experimental LLT dielectric permittivity data at $225\ \kelvin$. Both methods are capable of yielding well defined unique distributions for a given data set. 

                                

\begin{thebibliography}{24}
\expandafter\ifx\csname natexlab\endcsname\relax\def\natexlab#1{#1}\fi
\expandafter\ifx\csname bibnamefont\endcsname\relax
  \def\bibnamefont#1{#1}\fi
\expandafter\ifx\csname bibfnamefont\endcsname\relax
  \def\bibfnamefont#1{#1}\fi
\expandafter\ifx\csname citenamefont\endcsname\relax
  \def\citenamefont#1{#1}\fi
\expandafter\ifx\csname url\endcsname\relax
  \def\url#1{\texttt{#1}}\fi
\expandafter\ifx\csname urlprefix\endcsname\relax\def\urlprefix{URL }\fi
\providecommand{\bibinfo}[2]{#2}
\providecommand{\eprint}[2][]{\url{#2}}

\bibitem[{\citenamefont{Jonscher}(1983)}]{Jonscher1983}
\bibinfo{author}{\bibfnamefont{A.~K.} \bibnamefont{Jonscher}},
  \emph{\bibinfo{title}{Dielectric Relaxation in Solids}}
  (\bibinfo{publisher}{London: Chelsea Dielectric}, \bibinfo{address}{London},
  \bibinfo{year}{1983}); 
\bibinfo{editor}{\bibfnamefont{J.~R.} \bibnamefont{Macdonald}}, ed.,
  \emph{\bibinfo{title}{Impedance Spectroscopy}} (\bibinfo{publisher}{John
  Wiley \& Sons}, \bibinfo{address}{New York}, \bibinfo{year}{1987}); 
\bibinfo{editor}{\bibfnamefont{F.}~\bibnamefont{Kremer}} \bibnamefont{and}
  \bibinfo{editor}{\bibfnamefont{A.}~\bibnamefont{Sch{\"o}nhals}}, eds.,
  \emph{\bibinfo{title}{Broadband Dielectric Spectroscopy}}
  (\bibinfo{publisher}{Springer-Verlag}, \bibinfo{address}{Berlin},
  \bibinfo{year}{2003}).

\bibitem[{\citenamefont{Kohlrausch}(1854)}]{Kohl}
\bibinfo{author}{\bibfnamefont{R.}~\bibnamefont{Kohlrausch}},
  \bibinfo{journal}{Prog Ann der Phys Chem}
  \textbf{\bibinfo{volume}{91}}, \bibinfo{pages}{179} (\bibinfo{year}{1854});
\bibinfo{author}{\bibfnamefont{G.}~\bibnamefont{Williams}} \bibnamefont{and}
  \bibinfo{author}{\bibfnamefont{D.~C.} \bibnamefont{Watts}},
  \bibinfo{journal}{Trans Faraday Soc}
  \textbf{\bibinfo{volume}{66}}, \bibinfo{pages}{80} (\bibinfo{year}{1970});
\bibinfo{author}{\bibfnamefont{J.~R.} \bibnamefont{Macdonald}},
  \bibinfo{journal}{J Appl Phys}
  \textbf{\bibinfo{volume}{82}}(\bibinfo{number}{8}), \bibinfo{pages}{3962}
  (\bibinfo{year}{1997}).

\bibitem[{\citenamefont{McCrum et~al.}(1967)\citenamefont{McCrum, Read, and
  Williams}}]{McCrum}
\bibinfo{author}{\bibfnamefont{N.~G.} \bibnamefont{McCrum}},
  \bibinfo{author}{\bibfnamefont{B.~E.} \bibnamefont{Read}}, \bibnamefont{and}
  \bibinfo{author}{\bibfnamefont{G.}~\bibnamefont{Williams}},
  \emph{\bibinfo{title}{Anelastic and Dielectric Effects in Polymeric Solids}}
  (\bibinfo{publisher}{John Wiley \& Sons Ltd.}, \bibinfo{address}{London},
  \bibinfo{year}{1967}), \bibinfo{edition}{dover} ed.

\bibitem[{\citenamefont{Havriliak and Negami}(1966)}]{HN}
\bibinfo{author}{\bibfnamefont{S.}~\bibnamefont{Havriliak}} \bibnamefont{and}
  \bibinfo{author}{\bibfnamefont{S.}~\bibnamefont{Negami}},
  \bibinfo{journal}{J Polym Sci: Part C}
  \textbf{\bibinfo{volume}{14}}, \bibinfo{pages}{99} (\bibinfo{year}{1966}).

\bibitem[{\citenamefont{B{\"o}ttcher and Bordewijk}(1996)}]{BottcherDRT}
\bibinfo{author}{\bibfnamefont{C.~J.~F.} \bibnamefont{B{\"o}ttcher}}
  \bibnamefont{and}
  \bibinfo{author}{\bibfnamefont{P.}~\bibnamefont{Bordewijk}},
  \emph{\bibinfo{title}{Theory of Electric Polarization}}
  (\bibinfo{publisher}{Elsevier}, \bibinfo{year}{1996}),
  chap.~\bibinfo{chapter}{IX}, pp. \bibinfo{pages}{45--137},
  \bibinfo{edition}{third impression} ed.

\bibitem[{\citenamefont{Macdonald}(1999)}]{RossBJP}
\bibinfo{author}{\bibfnamefont{J.~R.} \bibnamefont{Macdonald}},
  \bibinfo{journal}{Braz J Phys}
  \textbf{\bibinfo{volume}{29}}(\bibinfo{number}{2}), \bibinfo{pages}{332}
  (\bibinfo{year}{1999}).

\bibitem[{\citenamefont{Macdonald}(1995)}]{macdonald:6241}
\bibinfo{author}{\bibfnamefont{J.~R.} \bibnamefont{Macdonald}},
  \bibinfo{journal}{J Chem Phys}
  \textbf{\bibinfo{volume}{102}}(\bibinfo{number}{15})
  (\bibinfo{year}{1995}{\natexlab{b}});
\bibinfo{author}{\bibfnamefont{J.~R.} \bibnamefont{Macdonald}},
  \bibinfo{journal}{J Comp Phys}
  \textbf{\bibinfo{volume}{157}}, \bibinfo{pages}{280}
  (\bibinfo{year}{2000}{\natexlab{a}});
\bibinfo{author}{\bibfnamefont{J.~R.} \bibnamefont{Macdonald}},
  \bibinfo{journal}{Inverse Probl} \textbf{\bibinfo{volume}{16}},
  \bibinfo{pages}{1561} (\bibinfo{year}{2000}{\natexlab{b}}).

\bibitem[{\citenamefont{Tuncer}(2000)}]{TuncerLicDRT}
\bibinfo{author}{\bibfnamefont{E.}~\bibnamefont{Tuncer}},
\bibinfo{type}{Lic 
  thesis--Tech rep} \bibinfo{number}{338~L}, 
\bibinfo{address}{Chalmers Univ of Technol, Gothenburg, Sweden} (\bibinfo{year}{2000}), \bibinfo{note}{ch. 5
  p63-83};
\bibinfo{author}{\bibfnamefont{E.}~\bibnamefont{Tuncer}} \bibnamefont{and}
  \bibinfo{author}{\bibfnamefont{S.~M.} \bibnamefont{Guba{\'n}ski}},
  \bibinfo{journal}{IEEE T Dielect El In}
  \textbf{\bibinfo{volume}{8}}, \bibinfo{pages}{310} (\bibinfo{year}{2001});
\bibinfo{author}{\bibfnamefont{E.}~\bibnamefont{Tuncer}},
  \bibinfo{author}{\bibfnamefont{M.}~\bibnamefont{Furlani}}, \bibnamefont{and}
  \bibinfo{author}{\bibfnamefont{B.-E.} \bibnamefont{Mellander}},
  \bibinfo{journal}{J Appl Phys}
  \textbf{\bibinfo{volume}{95}}(\bibinfo{number}{6}), \bibinfo{pages}{3131}
  (\bibinfo{year}{2004}).


\bibitem[{\citenamefont{Dias}(1996)}]{Dias}
\bibinfo{author}{\bibfnamefont{H.}~\bibnamefont{Kliem}},
  \bibinfo{author}{\bibfnamefont{P.}~\bibnamefont{Fuhrmann}}, \bibnamefont{and}
  \bibinfo{author}{\bibfnamefont{G.}~\bibnamefont{Arlt}},
  \bibinfo{journal}{IEEE T Elect Insulation}
  \textbf{\bibinfo{volume}{23}}(\bibinfo{number}{6}), \bibinfo{pages}{919}
  (\bibinfo{year}{1988});
\bibinfo{author}{\bibfnamefont{K.}~\bibnamefont{Tittelbach-Helmrich}},
  \bibinfo{journal}{Meas Sci Technol}
  \textbf{\bibinfo{volume}{4}}, \bibinfo{pages}{1323} (\bibinfo{year}{1993});
\bibinfo{author}{\bibfnamefont{F.~D.} \bibnamefont{Morgan}} \bibnamefont{and}
  \bibinfo{author}{\bibfnamefont{D.~P.} \bibnamefont{Lesmes}},
  \bibinfo{journal}{J Chem Phys}
  \textbf{\bibinfo{volume}{100}}(\bibinfo{number}{1}), \bibinfo{pages}{671}
  (\bibinfo{year}{1994});
\bibinfo{author}{\bibfnamefont{H.}~\bibnamefont{Sch{\"a}fer}},
  \bibinfo{author}{\bibfnamefont{E.}~\bibnamefont{Sternin}},
  \bibinfo{author}{\bibfnamefont{R.}~\bibnamefont{Stannarius}},
  \bibinfo{author}{\bibfnamefont{M.}~\bibnamefont{Arndt}}, \bibnamefont{and}
  \bibinfo{author}{\bibfnamefont{F.}~\bibnamefont{Kremer}},
  \bibinfo{journal}{Phys Rev Lett}
  \textbf{\bibinfo{volume}{76}}(\bibinfo{number}{12}), \bibinfo{pages}{2177}
  (\bibinfo{year}{1996});
\bibinfo{author}{\bibfnamefont{C.~J.} \bibnamefont{Dias}},
  \bibinfo{journal}{Phys Rev B}
  \textbf{\bibinfo{volume}{53}}(\bibinfo{number}{21}), \bibinfo{pages}{14212}
  (\bibinfo{year}{1996});
\bibinfo{author}{\bibfnamefont{M.}~\bibnamefont{Carmona}},
  \bibinfo{author}{\bibfnamefont{S.}~\bibnamefont{Marco}},
  \bibinfo{author}{\bibfnamefont{J.}~\bibnamefont{Palac\'{i}n}},
  \bibnamefont{and} \bibinfo{author}{\bibfnamefont{J.}~\bibnamefont{Samitier}},
  \bibinfo{journal}{IEEE T Comp Packag Technol}
  \textbf{\bibinfo{volume}{22}}(\bibinfo{number}{2}), \bibinfo{pages}{238}
  (\bibinfo{year}{1999});
\bibinfo{author}{\bibfnamefont{S.~M. F.~D.} \bibnamefont{Mustapha}}
  \bibnamefont{and} \bibinfo{author}{\bibfnamefont{T.~N.}
  \bibnamefont{Phillips}}, \bibinfo{journal}{J Phys D: Appl
  Phys} \textbf{\bibinfo{volume}{33}}, \bibinfo{pages}{1219}
  (\bibinfo{year}{2000}).

\bibitem[{\citenamefont{Stoneham}(1969)}]{Stoneham1969}
\bibinfo{author}{\bibfnamefont{A.~M.} \bibnamefont{Stoneham}},
  \bibinfo{journal}{Rev Mod Phys}
  \textbf{\bibinfo{volume}{41}}(\bibinfo{number}{1}), \bibinfo{pages}{82}
  (\bibinfo{year}{1969})

\bibitem[{\citenamefont{Schiener et~al.}(1996)\citenamefont{Schiener,
  B{\"o}hmer, Loidl, and Chamberlin}}]{BoehmerScience}
\bibinfo{author}{\bibfnamefont{B.}~\bibnamefont{Schiener}},
  \bibinfo{author}{\bibfnamefont{R.}~\bibnamefont{B{\"o}hmer}},
  \bibinfo{author}{\bibfnamefont{A.}~\bibnamefont{Loidl}}, \bibnamefont{and}
  \bibinfo{author}{\bibfnamefont{R.~V.} \bibnamefont{Chamberlin}},
  \bibinfo{journal}{Science} \textbf{\bibinfo{volume}{274}},
  \bibinfo{pages}{752} (\bibinfo{year}{1996});
\bibinfo{author}{\bibfnamefont{R.}~\bibnamefont{B{\"o}hmer}} \bibnamefont{and}
  \bibinfo{author}{\bibfnamefont{G.} \bibnamefont{Diezemann}}, in
  \emph{\bibinfo{booktitle}{Broadband Dielectric Spectroscopy}}, edited by
  \bibinfo{editor}{\bibfnamefont{F.}~\bibnamefont{Kremer}} \bibnamefont{and}
  \bibinfo{editor}{\bibfnamefont{A.}~\bibnamefont{Sch{\"o}nhals}}
  (\bibinfo{publisher}{Springer-Verlag}, \bibinfo{address}{Berlin},
  \bibinfo{year}{2003}), chap.~\bibinfo{chapter}{14}.

\bibitem[{\citenamefont{Leon et~al.}(1998)\citenamefont{Le{\'o}n, Santamaria,
  Paris, Sanz, Ibarra, and V{\'a}rez}}]{LeonJNCS}
\bibinfo{author}{\bibfnamefont{C.}~\bibnamefont{Le{\'o}n}},
  \bibinfo{author}{\bibfnamefont{J.}~\bibnamefont{Santamaria}},
  \bibinfo{author}{\bibfnamefont{M.~A.} \bibnamefont{Paris}},
  \bibinfo{author}{\bibfnamefont{J.}~\bibnamefont{Sanz}},
  \bibinfo{author}{\bibfnamefont{J.}~\bibnamefont{Ibarra}}, \bibnamefont{and}
  \bibinfo{author}{\bibfnamefont{A.}~\bibnamefont{V{\'a}rez}},
  \bibinfo{journal}{J Non-Cryst Solids}
  \textbf{\bibinfo{volume}{235-237}}, \bibinfo{pages}{753}
  (\bibinfo{year}{1998}).

\bibitem[{\citenamefont{Macdonald}(2004)}]{Macdonald2004PRB}
\bibinfo{author}{\bibfnamefont{J.~R.} \bibnamefont{Macdonald}}
  (\bibinfo{year}{2004}), \bibinfo{note}{submitted to Phys Rev B}.

\bibitem[{\citenamefont{Macdonald and Jr.}(1987)}]{LEVM}
\bibinfo{author}{\bibfnamefont{J.~R.} \bibnamefont{Macdonald}}
  \bibnamefont{and} \bibinfo{author}{\bibfnamefont{L.~D.~Potter}
  \bibnamefont{Jr.}}, \bibinfo{journal}{Solid State Ionics}
  \textbf{\bibinfo{volume}{24}}(\bibinfo{number}{1}), \bibinfo{pages}{61}
  (\bibinfo{year}{1987}), \bibinfo{note}{the latest version of the LEVM fitting program, V. 8.0, may be obtained at no cost from http//www.physics.unc.edu/{$\sim$}macd/ where more details about the program appear. An extensive manual, source code, and executable code are included.}

\bibitem[{\citenamefont{Butkov}(1968)}]{Butkov}
\bibinfo{author}{\bibfnamefont{E.}~\bibnamefont{Butkov}},
  \emph{\bibinfo{title}{Mathematical Physics}}, Addison-Wesley Series in
  Advanced Physics (\bibinfo{publisher}{Addison-Wesley Publishing Company},
  \bibinfo{address}{Menlo Park}, \bibinfo{year}{1968}).

\bibitem[{\citenamefont{Wolpert et~al.}(2003)\citenamefont{Wolpert, Ickstadt,
  and Hansen}}]{Wolpert}
\bibinfo{author}{\bibfnamefont{R.~L.} \bibnamefont{Wolpert}},
  \bibinfo{author}{\bibfnamefont{K.}~\bibnamefont{Ickstadt}}, \bibnamefont{and}
  \bibinfo{author}{\bibfnamefont{M.~B.} \bibnamefont{Hansen}}, in
  \emph{\bibinfo{booktitle}{Bayesian Statistics 7}}, edited by
  \bibinfo{editor}{\bibfnamefont{M.~J.} \bibnamefont{Bernardo}},
  \bibinfo{editor}{\bibfnamefont{M.~J.} \bibnamefont{Bayarri}},
  \bibinfo{editor}{\bibfnamefont{J.~O.} \bibnamefont{Berger}},
  \bibinfo{editor}{\bibfnamefont{A.~P.} \bibnamefont{Dawid}},
  \bibinfo{editor}{\bibfnamefont{D.}~\bibnamefont{Heckerman}},
  \bibinfo{editor}{\bibfnamefont{A.~F.~M.} \bibnamefont{Smith}},
  \bibnamefont{and} \bibinfo{editor}{\bibfnamefont{M.}~\bibnamefont{West}}
  (\bibinfo{address}{Oxford}, \bibinfo{year}{2003}), pp.
  \bibinfo{pages}{403--418};
\bibinfo{author}{\bibfnamefont{R.~L.} \bibnamefont{Wolpert}} \bibnamefont{and}
  \bibinfo{author}{\bibfnamefont{K.}~\bibnamefont{Ickstadt}},
  \bibinfo{journal}{Inverse Probl} \textbf{\bibinfo{volume}{20}},
  \bibinfo{pages}{1759} (\bibinfo{year}{2004}).


\bibitem[{\citenamefont{Macdonald}(2002)}]{RossSSI2002}
\bibinfo{author}{\bibfnamefont{J.~R.} \bibnamefont{Macdonald}},
  \bibinfo{journal}{J Non-Cryst Solids}
  \textbf{\bibinfo{volume}{210}}, \bibinfo{pages}{70}
  (\bibinfo{year}{1997}{\natexlab{a}});
\bibinfo{author}{\bibfnamefont{J.~R.} \bibnamefont{Macdonald}},
  \bibinfo{journal}{J Non-Cryst Solids}
  \textbf{\bibinfo{volume}{212}}, \bibinfo{pages}{95}
  (\bibinfo{year}{1997}{\natexlab{b}});
\bibinfo{author}{\bibfnamefont{J.~R.} \bibnamefont{Macdonald}},
  \bibinfo{journal}{Solid State Ionics} \textbf{\bibinfo{volume}{150}},
  \bibinfo{pages}{263} (\bibinfo{year}{2002}).


\bibitem[{\citenamefont{Debye}(1945)}]{Debye1945}
\bibinfo{author}{\bibfnamefont{P.}~\bibnamefont{Debye}},
  \emph{\bibinfo{title}{Polar Molecules}} (\bibinfo{publisher}{Dover
  Publications}, \bibinfo{address}{New York}, \bibinfo{year}{1945}).

\end{thebibliography}
\end{document}